\documentclass[a4paper,12pt]{article}
\pdfoutput=1
\usepackage{graphicx, rotating,colortbl,amssymb}
\usepackage{hyperref}
\usepackage{slashed}
\usepackage{ifpdf}
\usepackage{charter}
\usepackage[dvipsnames]{xcolor}

\def\bea{\begin{eqnarray}}
\def\eea{\end{eqnarray}}

\ifx\pdfoutput\undefined
\usepackage[dvips,bookmarks=false]{hyperref}	
\else
\usepackage{hyperref}	
\fi
\hypersetup{colorlinks,bookmarksopen,bookmarksnumbered,citecolor=verdes,
linkcolor=blus,pdfstartview=FitH,urlcolor=rossos}
\def\hhref#1{\href{http://arxiv.org/abs/#1}{#1}} 

\usepackage{amsfonts}
\usepackage{amsmath}
\usepackage{slashed}

\newcommand{\beq}{\begin{equation}}
\newcommand{\eeq}{\end{equation}}
\newcommand{\fig}[1]{~\ref{fig:#1}}

\newcount\Mac  \Mac=1  
\newcommand{\ifMac}[2]{\ifnum\Mac=1 #1 \else #2 \fi}
\def\putps(#1,#2)(#3,#4)#5#6{\ifnum\Mac=1 \put(#1,#2){\special{picture #5}}
\else  \put(#3,#4){\includegraphics{#6}} \fi}

\newcommand{\One}{\hbox{1\kern-.24em I}}

\newcommand{\GeV}{\,{\rm GeV}}
\newcommand{\TeV}{\,{\rm TeV}}

\newcommand{\eq}[1]{~{\rm (\ref{eq:#1})}}

\newcommand{\lascia}[1]{}
\makeatletter
\def\art{\@ifnextchar[{\eart}{\oart}}
\def\eart[#1]#2#3#4#5#6{{\rm #2}, {#3 #4} {\rm (#6) #5} [arXiv:{\hhref{#1}}]}
\def\hepart[#1]#2{{\rm #2, arXiv:\hhref{#1}}}
\newcommand{\oart}[5]{{\rm #1}, {#2 #3} {\rm (#5) #4}}

\newcounter{alphaequation}[equation]
\def\thealphaequation{\theequation\hbox to
0.6em{\hfil\alph{alphaequation}\hfil}}
\def\eqnsystem#1{
\def\@eqnnum{{\rm (\thealphaequation)}}
\def\@@eqncr{\let\@tempa\relax \ifcase\@eqcnt \def\@tempa{& & &} \or
  \def\@tempa{& &}\or \def\@tempa{&}\fi\@tempa
  \if@eqnsw\@eqnnum\refstepcounter{alphaequation}\fi
\global\@eqnswtrue\global\@eqcnt=0\cr}
\refstepcounter{equation} \let\@currentlabel\theequation \def\@tempb{#1}
\ifx\@tempb\empty\else\label{#1}\fi
\refstepcounter{alphaequation}
\let\@currentlabel\thealphaequation
\global\@eqnswtrue\global\@eqcnt=0 \tabskip\@centering\let\\=\@eqncr
$$\halign to \displaywidth\bgroup \@eqnsel\hskip\@centering
$\displaystyle\tabskip\z@{##}$&\global\@eqcnt\@ne
\hskip2\arraycolsep\hfil${##}$\hfil& \global\@eqcnt\tw@\hskip2\arraycolsep
$\displaystyle\tabskip\z@{##}$\hfil
\tabskip\@centering&\llap{##}\tabskip\z@\cr}

\def\endeqnsystem{\@@eqncr\egroup$$\global\@ignoretrue} \makeatother

\def\SU{{\rm SU}}

\def\circa#1{\,\raise.3ex\hbox{$#1$\kern-.75em\lower1ex\hbox{$\sim$}}\,}

\usepackage{multicol}
\usepackage{color}
\definecolor{rosso}{cmyk}{0,1,1,0.4}
\definecolor{rossos}{cmyk}{0,1,1,0.55}
\definecolor{rossoc}{cmyk}{0,1,1,0.2}
\definecolor{blu}{cmyk}{1,1,0,0.3}
\definecolor{blus}{cmyk}{1,1,0,0.6}
\definecolor{bluc}{cmyk}{1,1,0,0.1}
\definecolor{verde}{cmyk}{0.92,0,0.59,0.25}
\definecolor{verdec}{cmyk}{0.92,0,0.59,0.15}
\definecolor{verdes}{cmyk}{0.92,0,0.59,0.4}
\definecolor{grigio}{cmyk}{0,0,0,0.07}
\definecolor{rosa}{cmyk}{0,0.1,0.1,0.02}
\definecolor{rosino}{cmyk}{0,0.05,0.05,0.02}
\definecolor{rosas}{cmyk}{0,0.3,0.25,0.05}
\definecolor{celeste}{cmyk}{0.1,0,0,0.02}
\definecolor{giallino}{cmyk}{0,0,0.4,0.02}
\definecolor{rosso}{cmyk}{0,1,1,0.4}
\definecolor{rossos}{cmyk}{0,1,1,0.55}
\definecolor{rossoc}{cmyk}{0,1,1,0.2}
\definecolor{blu}{cmyk}{1,1,0,0.3}
\definecolor{bluc}{cmyk}{1,1,0,0.1}
\definecolor{blucc}{cmyk}{0.7,0.5,0,0}
\definecolor{viola}{cmyk}{0,1,0,0.6}
\definecolor{viola2}{cmyk}{0,1,0.2,0.6}
\definecolor{verde}{cmyk}{0.92,0,0.59,0.25}
\definecolor{verdec}{cmyk}{0.92,0,0.59,0.15}
\definecolor{verdes}{cmyk}{0.92,0,0.59,0.4}
\definecolor{verdino}{cmyk}{0.12,0,0.09,0.05}
\definecolor{giallo}{cmyk}{0,0,1,0}
\definecolor{gialloverde}{cmyk}{0.44,0,0.74,0}

\font\tenrsfs=rsfs10 at 12pt

\font\sevenrsfs=rsfs7
\font\fiversfs=rsfs5
\newfam\rsfsfam
\textfont\rsfsfam=\tenrsfs
\scriptfont\rsfsfam=\sevenrsfs
\scriptscriptfont\rsfsfam=\fiversfs
\def\mathscr#1{{\fam\rsfsfam\relax#1}}

\begin{document}
\color{black}
\begin{center}
{\Huge\bf\color{Magenta}Dynamical generation of the\\{}~weak  and Dark Matter scale
}
\bigskip\color{black}\vspace{0.3cm} \\[3mm]
{{\large\bf  Thomas Hambye$^{a}$ {\rm and}  Alessandro Strumia$^{b,c}$}
} \\[5mm]
{\it  (a)  Service de Physique Th\'eorique,
 Universit\'e Libre de Bruxelles, Brussels, Belgium}\\
{\it  (b) Dipartimento di Fisica dell'Universit{\`a} di Pisa and Istituto Nazionale Fisica Nucleare, Italia}\\
{\it  (c) National Institute of Chemical Physics and Biophysics, Tallinn, Estonia}\\
\end{center}
\bigskip

\centerline{\large\bf\color{blus} Abstract}
\begin{quote} 
Assuming that naturalness should be modified by ignoring
quadratic divergences, we propose a simple extension of the Standard Model where
the weak scale is dynamically generated together with an automatically stable vector.
Identifying it as thermal Dark Matter, the model has one free parameter.
It predicts one  extra scalar, detectable at colliders, 
which triggers
a first-order dark/electroweak cosmological phase transition 
with production of gravitational waves.
Vacuum stability holds up to the Planck scale.
 \color{black}
\end{quote}


\section{Introduction}
The discovery of the SM scalar~\cite{Higgsdata} together with
 negative results of searches  for supersymmetry and for other solutions to the usual hierarchy problem~\cite{thooft}
invite us to explore the idea that this paradigm needs to be abandoned or reformulated.
One possibility is that naturalness still holds but in a modified version, 
namely under the assumption that the unknown cut-off at Planckian scales
has the property that quadratic divergences vanish, like in dimensional regularisation.
Such modified  `finite naturalness'  was discussed in~\cite{FPS}, showing that it is satisfied by the Standard Model 
and that new physics models motivated by data (about
neutrino masses, Dark Matter, QCD $\theta$ problem, inflation) can satisfy it.

\smallskip

In this paper we address what could be the dynamical origin of the weak scale and of the Dark Matter scale in the context of
`finite naturalness'. 
Although not logically necessary, the extra hypothesis that mass terms are absent from the fundamental Lagrangian
may be conceptually more appealing than just putting  by hand small masses of the order of the  electroweak scale.
We adopt the following three guidelines:
\begin{enumerate}
\item We assume that the SM and DM particles have no mass terms in the fundamental Lagrangian,
and that their masses arise from some dynamical mechanism.

\item
We assume that the extended theory has the same automatic properties of the SM supported by data:
accidental conservation of lepton and baryon number, of lepton flavour, etc.

\item We assume that DM stability is one more automatic consequence of the theory.
\end{enumerate}
The resulting model is presented in section~\ref{model} and its phenomenology is explored in section~\ref{phono}.
In section~\ref{concl} we conclude.

\section{The model}\label{model}
One simple model is obtained by merging previously proposed ideas that possess some of the properties 1,2,3: 
ref.~\cite{CW,prev} for
dynamical generation of the weak  scale
(see also~\cite{raidal} for related ideas), and ref.~\cite{VDM,MDM} for automatic (accidental) DM stability. 

\medskip

The model has gauge group ${\rm U}(1)_Y\otimes\SU(2)_L\otimes\SU(3)_c\otimes\SU(2)_X$, namely the SM gauge group with an extra $\SU(2)_X$.  The field content is just given by the SM fields (singlets under $\SU(2)_X$) plus a scalar $S$,
doublet  under the extra $\SU(2)_X$ and neutral under the SM gauge group.
The Lagrangian of the model is just the most general one, omitting the mass terms for the SM scalar doublet $H$ (``Higgs" for short) and for the
scalar doublet $S$, because we want to dynamically generate the weak and DM scales.
Consequently, the scalar potential of the theory is:
\beq V = \lambda_H |H|^4 -\lambda_{HS} |H S|^2 + \lambda_S |S|^4.\label{eq:V}\eeq
We now show how it can lead to dynamical symmetry breaking down to ${\rm U}(1)_{\rm em}\otimes \SU(3)_c$
such that, in unitary gauge, the scalar doublets $H$ and $S$ can be expanded in components $h$ and $s$ as
\beq H(x) = \frac{1}{\sqrt{2}}
\begin{pmatrix}
0\cr v+h(x)
\end{pmatrix},\qquad
S(x) = \frac{1}{\sqrt{2}} \begin{pmatrix}
0\cr
w+s(x)
\end{pmatrix},\eeq
where $v\approx 246\GeV$ is the usual Higgs vacuum expectation value (vev),
and $w$ is the vev that completely breaks $\SU(2)_X$
giving equal masses $M_X = g_X w/2$ to all $\SU(2)_X$ vectors.
Symmetry breaking  happens when \cite{Sher:1988mj}
\beq 4\lambda_H \lambda_S-  \lambda_{HS}^2<0,
\label{eq:SB}\eeq
a condition that can be dynamically verified at low energy because quantum corrections make $\lambda_S$ smaller at low energy,
as described by the beta function
\beq  \beta_{\lambda_S} \equiv \frac{d\lambda _S}{d\ln\mu} = \frac{1}{(4\pi)^2}\bigg[
\frac{9 g_X^4}{8}-9 g_X^2 \lambda _{{S}}+2 \lambda _{{HS}}^2+24 \lambda _{{S}}^2 
\bigg].\eeq
Unlike in the $\beta$ function of $\lambda_H$, there is no negative Yukawa contribution:  $\beta_{\lambda_S}$ is definite positive and the gauge term
makes $\lambda_S$ negative at low energy. 
Thereby the dynamically generated hierarchy between $v\sim w$ and the Planck scale is 
exponentially large, of order
$e^{\lambda_S/\beta_{\lambda_S}}$.

While the analysis of the full one-loop potential is somehow involved, 
a simple analytic approximation holds in the
limit of small $\lambda_{HS}$ (which will be phenomenologically justified a posteriori).
In this limit the instability condition of eq.\eq{SB} can be approximated as $\lambda_S<0$ and
the potential at one loop order can be approximated by inserting a running
$\lambda_S$ in the tree-level potential of eq.\eq{V}:
\beq
\lambda_S \simeq  \beta_{\lambda_S} \ln s/s_*  ,\eeq
where 
$s_*$ is the critical scale below which $\lambda_S$ becomes negative.
The use of $s_*$ is not an approximation but a convenient parameterisation.
Given that around $s\sim s_*$ the typical size of $\lambda_S$ 
 is $\beta_{\lambda_S}$,
``small $\lambda_{HS}$'' in eq.\eq{SB} precisely means $\lambda_{HS}^2\ll \lambda_H\beta_{\lambda_S}$.
In this approximation,
the potential is minimised as
\beq
v \simeq w \sqrt{\frac{\lambda_{HS}}{2\lambda_H}},\qquad
w \simeq s_* e^{-1/4}.\eeq
The scalar mass matrix at the minimum in the ($h,s$) basis is:
\beq v^2
\begin{pmatrix}
2\lambda_H & - \sqrt{2\lambda_H\lambda_{HS}} \cr
- \sqrt{2\lambda_H\lambda_{HS}}  & \lambda_{HS} + 2\beta_{\lambda_S} {\lambda_H}/{\lambda_{HS}} 
\end{pmatrix}\,.
\eeq
Its eigenvalues 
$m_1$ and $m_2$ are
\beq m_1^2 \simeq  2 v^2 \lambda,\qquad
m_2^2 \simeq  v^2  \frac{2\beta_{\lambda_S}\lambda_H}{\lambda_{HS}} ,
\label{eq:eigen}
\eeq
where $\lambda \simeq \lambda_H -\lambda_{HS}^2/\beta_{\lambda_S}$,
having included the next-to-leading term in the small $\lambda_{HS}$ approximation
(as discussed later, this small effect helps in keeping $\lambda_H$ and the whole potential stable
up to large Planckian field values~\cite{SS}).
The first state, $h_1$ can be identified with the Higgs boson with mass $m_1 \approx 125.6\GeV$~\cite{Higgsdata}.
The mixing angle, defined by $h_1 = h \cos\alpha + s \sin\alpha$, is given by
\beq \sin 2\alpha =\frac{v^2 \sqrt{8\lambda_H \lambda_{HS}}}{m_2^2-m_1^2}\qquad
\hbox{i.e.}\qquad
\alpha \stackrel{m_1\ll m_2}{\simeq} \frac{\lambda_{HS}^{3/2}}{\sqrt{2\lambda_H}\beta_{\lambda_S}}.
\eeq
Due to mixing, the extra state $h_2=s\cos\alpha- h \sin\alpha$ inherits the couplings to SM particles of the Higgs boson $h$, rescaled by the factor $\sin\alpha$.
Note that   $m_1^2 $ can be rewritten as $m_1^2 \simeq w^2 \lambda_{HS}$, showing explicitly that the
SM scalar boson mass is induced by the $\lambda_{HS}$ portal, proportionally to the $\SU(2)_X$ gauge symmetry breaking scale $w$. 
Electroweak symmetry breaking does not need a large value for the portal coupling $\lambda_{HS}$, provided that $w$ is large enough.\footnote{This differs from the DM driven EWSB mechanism proposed in ref.~\cite{HT} in the framework of the inert Higgs doublet model (see also \cite{EspinosaQuiros,EspinosaKNoQuiros} with singlet scalars). In these models
only the SM scalar field gets a vacuum expectation value, so that in order to compensate
the negative top Yukawa coupling contribution to $\beta_{\lambda_H}$, EWSB requires either quite large new quartic couplings (that
can become non-perturbative already at multi-TeV energies) or many extra scalars.}
Effectively $s$ acts as `the Higgs of the Higgs' and as `the Higgs of Dark Matter'.
Furthermore, the `Higgs of the $s$' is $s$ itself:
all scales are dynamically generated via dimensional transmutation.


\bigskip

As discussed in~\cite{VDM,VDM2,VDM3}, the SU(2)$_X$ vectors with mass $M_X $ are DM candidates,
automatically stable thanks to the analogous of the accidental custodial symmetry of the SM.\footnote{
DM vectors are a  triplet, and thereby cannot decay into the bosons, which are singlets under the dark custodial symmetry.
In this respect, it is important that the dark gauge group is $\SU(2)_X$.
In models that  employ instead an abelian U(1)$_X$ dark gauge group \cite{VDM,Farzan}, vectors are unstable unless
kinetic mixing with U(1)$_Y$ is forbidden.
In models with a  non-abelian group larger than $\SU(2)_X$ it is more difficult to find a scalar representation
with dimension just above the dimension of the gauge group, that  can fully break it (but possible if one makes extra assumptions on the structure
of the vacuum \cite{D'Eramo:2012rr}).}
Such symmetry can be violated only by non-renormalizable dimension-6 operators.
If suppressed by Planckian scales, such operators leave these hidden vector dark matter TeV-scale particles
stable enough on cosmological time-scales.

\section{Phenomenological analysis}\label{phono}
The DM thermal relic abundance and DM indirect signals require the computation of $\sigma v$, 
the non-relativistic DM annihilation cross section times the relative velocity $v$.
We compute them in the limit of small $\lambda_{HS}$, where  only 
$\SU(2)_X$ gauge interactions are relevant,
and the mixing angle $\alpha$ is small, such that DM dominantly annihilates into $s$.
The model gives rise to $VV\to ss$  annihilations and to
$VV\to Vs$ semi-annihilations \cite{VDM} (see also \cite{VDM2,VDM3,othersemi}).
Averaging over initial spin and gauge components we find:
\beq \sigma v_{\rm ann} =\frac{11g_X^2}{1728\pi w^2}
,\qquad
\sigma v_{\rm semi-ann} = \frac{g_X^2}{32\pi w^2}.
\eeq
The cosmological DM abundance is reproduced as a thermal relic for
$\sigma v_{\rm ann}+ \frac{1}{2}\sigma v_{\rm semi-ann}\approx 2.2~10^{-26}\,{\rm cm}^3/{\rm s}$, which means
\begin{equation}
g_X \simeq w/2.0\TeV .
\end{equation}

\bigskip

\begin{figure}[t]
\begin{center}
$$\includegraphics[height=7cm]{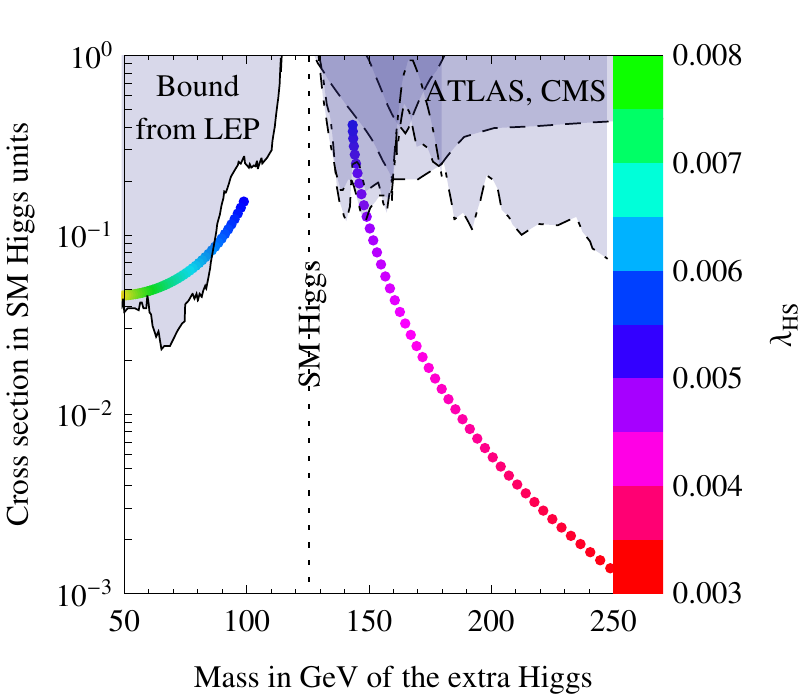}\qquad
\includegraphics[height=7cm]{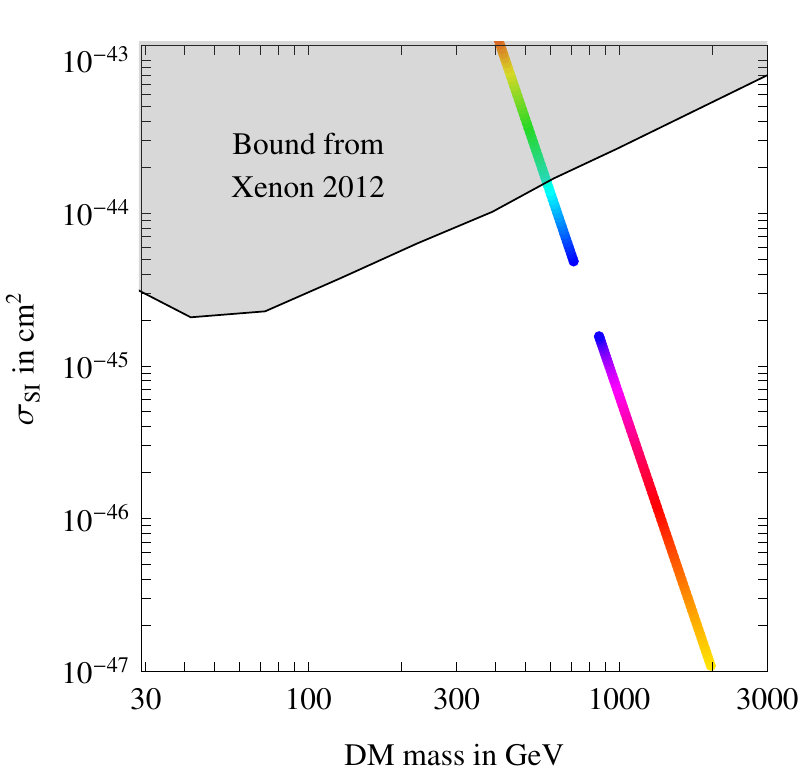}$$
\caption{\em   Predicted cross sections for the extra scalar boson  (left) and for DM direct detection  (right) as function of 
the only free parameter of the model $\lambda_{HS}$, varied as shown in the colour legend.
\label{fig:FS}}
\end{center}
\end{figure}

With this assumption, the model only has one free parameter:
out of the 4 parameters $\lambda,\lambda_{HS}, \lambda_S$ (or $s_*$) and $g_X$,
3 of them are fixed by the observed Higgs mass and vacuum expectation value, and by the cosmological DM abundance $\Omega_{\rm DM}$~\cite{Planck}.
We can view $g_X$ or $\lambda_{HS}$ as the only free parameter.
They are related as
\beq \lambda_{HS} \simeq 0.004/g_X^2\label{eq:rel}\eeq
The observables are predicted in terms of $g_X$ as
\beq
\begin{array}{rclrcl}
m_2 &\simeq& 165\GeV g_X^3,\qquad \qquad&
\alpha &\simeq& 0.07/g_X^7,\\
M_X &\simeq&  985\GeV g_X^2,&
\sigma_{\rm SI} 
&\simeq& 0.7~10^{-45}\,{\rm cm}^2\, \Big[1-(m_1^2/m_2^2)\Big]^2/g_X^{12} .
\end{array}\eeq
where $\sigma_{\rm SI}$ is the DM direct detection cross section, computed below.
These approximations hold in the limit of small $\lambda_{HS}$, which numerically means
$\lambda_{HS}\ll \sqrt{\lambda_H\lambda_S}\approx 0.015$ i.e.\ $g_X \gg 0.5$.
Fig.\fig{FS} shows the same predictions, computed numerically without making the small $\lambda_{HS}$ approximation.

\medskip

\subsubsection*{The Higgs of the Higgs}
Fig.\fig{FS}a shows the predictions for Higgs physics.  
We see that the new  scalar $h_2$ cannot have a mass in the range between 100 and 140 GeV.
Indeed, the two scalar states roughly have the same mass terms for $\lambda_{HS}\approx 0.005$;
however, due to the off-diagonal term in their mass matrix, the mass difference must be larger than
\beq | m_1-m_2|\circa{>} |m_{12}^2|/m_1 \approx m_1 \sqrt{\lambda_{HS}/2\lambda_H}\approx 17\GeV.\eeq
The extra state $h_2$ behaves as an extra Higgs boson with couplings rescaled by $\sin\alpha$.
This means that it is a narrow resonance even if heavier than 1 TeV.
For $m_2 < 2 m_1$ the extra scalar behaves as a Higgs-like state with production
cross section suppressed by $\sin^2\alpha$, while
for $m_2>2m_1$ the extra state also has a decay width into two Higgs,
\beq \Gamma (h_2\to h_1h_1) = \frac{\lambda_{HS}^2}{32\pi}\frac{w^2}{m_2^2}\sqrt{m_2^2-4 m_1^2},\eeq
which contributes to up $20\%$ to its total width, dominated by $h_2\to WW,ZZ,t\bar t$.
The shaded regions in fig.\fig{FS}a are excluded by LEP  (at small mass) and LHC (at large mass,
$h\to WW$ searches are plotted as dashed curves and $h\to ZZ$ searches as dot-dashed curves).
Future sensitivities are discussed in~\cite{Frere}.
Present experimental searches for $h\to ZZ$ 
and for $h\to\gamma\gamma$ show some 
(non statistically significant)
hint for an extra state at $m_2 \approx 143\GeV$~\cite{Higgsdata}.

The cross section for DM production at LHC (mediated by off-shell $h_1$ or $h_2$) can easily be negligibly small.

\medskip

\subsubsection*{Direct Dark Matter signals}
The Spin-Independent cross section for DM direct detection is~\cite{VDM}
\beq \sigma_{\rm SI} = \frac{m_N^4  f^2 }{16\pi v^2}
 \bigg(\frac{1}{m_1^2} - \frac{1}{m_2^2}\bigg)^2 g_X^2 \sin^2 2\alpha
\eeq
where $f\approx 0.295$ is the nucleon matrix element and $m_N$ is the nucleon mass.
Fig.\fig{FS}b shows the predictions for DM direct searches.  Present direct detection constraints imply
the bounds $\lambda_{HS}\lesssim 0.007$ (so that the 
approximation of small $\lambda_{HS}$ holds in the  
phenomenologically interesting region), $m_2\gtrsim 70$~GeV, $w\gtrsim1.5$~TeV, 
$M_X\gtrsim  560\GeV$,
$g_X\gtrsim 0.75$ and $\alpha\lesssim 0.5$. 
A value of $\lambda_{HS}=0.007$ is also disfavoured by LEP Higgs searches (fig.\fig{FS}a).
Future experiments should be able to probe smaller values of $\lambda_{HS}$ by improving the sensitivity to the extra Higgs cross section 
by $1-2$ orders of magnitudes 
and the sensitivity to $\sigma_{\rm SI}$ by $2-3$ orders of magnitude before 2020.

Signals disappear in the limit of small $\lambda_{HS}$.
However $\lambda_{HS}$ cannot be arbitrarily small, because eq.\eq{rel} would imply a non-perturbatively
large value for $g_X$.\footnote{If the $g_X$ coupling becomes non-perturbative at a scale $\Lambda_{X}$ 
larger than the scale where eq.~(\ref{eq:SB}) is satisfied, the dark scalars and vectors confine~\cite{VDM2}. In this case the $\lambda_{HS} S^\dagger S H^\dagger H $ term could also lead to EWSB by inducing a negative $\lambda_{HS} \Lambda^2_{X} H^\dagger H$ mass term \cite{VDM2}.}
In practice, values of $\lambda_{HS}$ below about $10^{-3}$ imply a large value of
the gauge coupling $g_X \gtrsim 2$  and correspond to
 $m_2\gtrsim 1.3$~TeV, $M_X\gtrsim 4$~TeV,  and $\alpha \lesssim 0.0005$.
 This is enough to render the signals too small to be observed in forthcoming experiments.

\smallskip

\subsubsection*{Indirect Dark Matter signals}
Concerning indirect DM signals, the model predicts DM annihilations into $ss$
with thermal cross section.  This determines the detectable astrophysical energy spectra of $e^+,\bar p,\gamma, \nu, \bar d$
as computed e.g.\ in~\cite{DMtools}.
Furthermore, DM semi-annihilations (around the center of the galaxy and of the Sun)
produce a flux of DM particles with $E=M_X$.
They could be detected by looking for hadronic showers
produced by DM scatterings with matter.
However the cross section 
$\sigma \sim g_X^4 m_N^2\sin^2 2\alpha/4\pi M_X^4$
is smaller than the cross-section for the analogous neutrino signals from DM annihilations.

\begin{figure}[t]
\begin{center}
$$\includegraphics[width=0.75\textwidth]{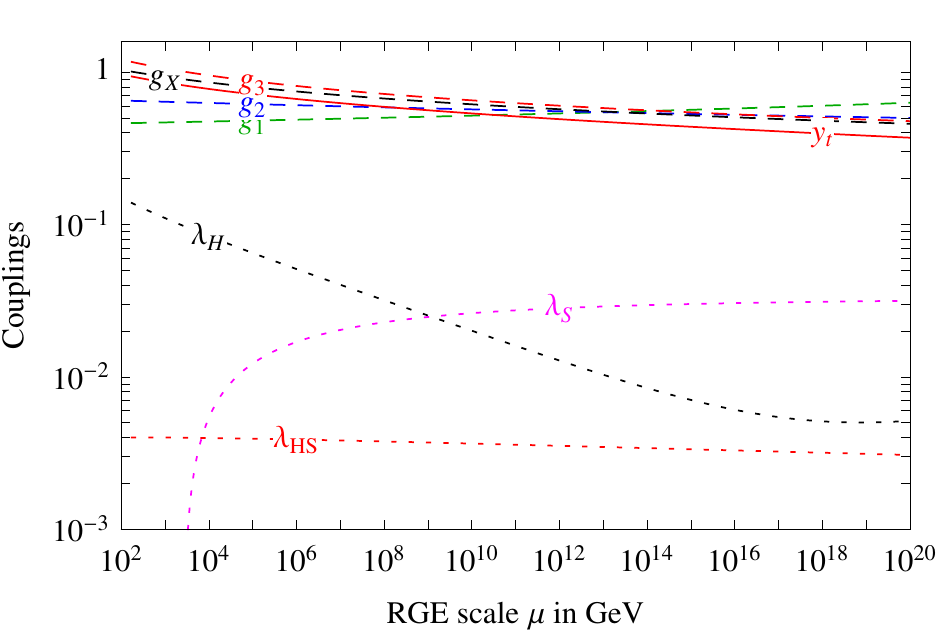}$$
\caption{\em Running of the model parameters up to the Planck scale for $g_X=1$.
\label{fig:run}}
\end{center}
\end{figure}

\subsubsection*{RGE and stability of the potential}
Having determined the weak-scale values of the parameters, 
we now explore how the model can be 
extrapolated up to large energies,  dynamically generating the weak scale.
The renormalisation group equations (RGE) for the model are those of the SM, 
with extra terms in the RGE for the quartic Higgs coupling
\beq (4\pi)^2\frac{d\lambda _{{H}}}{d\ln\mu} = (12 g_t^2-\frac{9 g_1^2}{5}-9 g_2^2) \lambda _{{H}}-6 g_t^4+\frac{27 g_1^4}{200}+\frac{9}{20} g_2^2 g_1^2+\frac{9 g_2^4}{8}+24 \lambda _{{H}}^2+2 \lambda _{{HS}}^2 \eeq
and supplemented by the RGE for the extra couplings:
\begin{eqnsystem}{sys:RGE}
(4\pi)^2\frac{dg_X}{d\ln\mu} &=& - \frac{43}{6}g_X^3
-\frac{1}{(4\pi)^2} \frac{259}{6}g_X^5+\cdots \\
(4\pi)^2\frac{d\lambda _{{HS}}}{d\ln\mu} &=& \lambda _{{HS}} \left(6 y_t^2-\frac{9 g_X^2}{2}-\frac{9 g_1^2}{10}-\frac{9 g_2^2}{2}+12 \lambda _{{H}}+12 \lambda _{{S}}\right)-4 \lambda _{{HS}}^2 \\
(4\pi)^2\frac{d\lambda _{{S}}}{d\ln\mu} &=& -9 g_X^2 \lambda _{{S}}+\frac{9 g_X^4}{8}+2 \lambda _{{HS}}^2+24 \lambda _{{S}}^2 
\end{eqnsystem}
Fig.\fig{run} shows the resulting running of the couplings of the model up to the Planck scale for $g_X=1$,
which corresponds to $\lambda_{HS} = 0.004$.
We notice that the model interprets the observed proximity of the QCD scale to the electroweak scale
as due to a proximity between the strong gauge coupling $g_3$ and the dark gauge coupling $g_X$.
Indeed, $g_3$ and $g_X$ happen to have not only similar values at the weak scale, but also
a numerically similar $\beta$ function, such that all gauge coupling roughly reach a common value at large energies. 
At low energy $g_X$ becomes large, of order one, triggering a negative $\lambda_S$ and consequently dynamically generating
the DM scale and the weak scale.

\medskip

\subsubsection*{Dark/electroweak phase transition}
The mechanism of dynamical scale generation
implies a negative value of the cosmological constant 
(barring meta-stable minima).
The contribution of the present model is $V_{\rm min} \simeq -w^4\beta_{\lambda_S}/16$.
Despite being suppressed by a one-loop factor, this contribution is larger by about 60 orders of
magnitude than the observed value.
Assuming that the cosmological constant problem is solved by a fine-tuning, we can proceed to study how the dark and electroweak 
phase transitions occur during the big-bang.

We recall that the SM predicts a second-order phase transition where the Higgs boson starts to obtain 
a vacuum expectation value $v(T)$ at temperatures below $T_c^{\rm SM}\approx 170\GeV$
and sphalerons decouple when $T_{\rm dec}^{\rm SM}\approx v(T_{\rm dec}^{\rm SM}) \approx140\GeV$~\cite{laine}.

Within the present model, using again the small $\lambda_{HS}$ approximation, the one-loop thermal correction to the potential is
\beq  V_T(s, h\approx 0) = \frac{9T^4}{2\pi^2} f(\frac{M_X}{T})+\frac{T}{4\pi}[M_X^3 - (M_X^2 + \Pi_X)^{3/2}]\label{eq:VT}
\eeq
where $f(r) =\int_0^\infty x^2 \ln(1-e^{-\sqrt{x^2+r^2}}) dx$ and $\Pi_X = 11 g_X^2 T^2/6$ is the thermal propagator
for the longitudinal $X$ component
which accounts for re-summation of higher order
daisy diagrams~\cite{EspinosaQuiros,EspinosaKNoQuiros}.
Eq.\eq{VT} predicts that $s$ and consequently $h$ acquire
a vacuum expectation value through a first-order phase transition.  The critical temperature at which the two
phases are degenerate is $T_c/M_X \simeq 0.37, 0.42,0.49,0.75$ for $g_X=0.75,1,1.2,1.5$ respectively.

\medskip

A cosmological first-order phase transition occurring at temperatures around the weak scale generates
gravitational waves at a potentially detectable level.

Their present peak frequency and energy density are~\cite{Grojean:2006bp,EspinosaKNoQuiros}
\beq f_{\rm peak}\approx 5\,{\rm mHz} \frac{\beta/H}{100} \frac{T_f}{1\TeV},\qquad
\Omega_{\rm peak} h^2 \approx 1.84~10^{-6} \kappa^2 \Big(\frac{\alpha}{1+\alpha}\Big)^2\frac{H^2}{\beta^2}\frac{v_b^3}{0.42+v_b^2}
\eeq
where $T_f\circa{<}T_c$ is the temperature at which the phase transition happens,
$\alpha$ is the energy fraction involved in the first-order phase transition, $H/\beta$ is the duration of the
phase transition in Hubble units, $v_b=(\sqrt{1/3}+\sqrt{\alpha^2+2\alpha/3})/(1+\alpha)$ is the wall velocity and $\kappa=(0.715 \alpha+\sqrt{8\alpha/243})/(1+0.715 \alpha)$ is the fraction of latent heat converted into gravitational waves.
$\alpha$ and $\beta$ are explicitly given by
\beq \alpha \equiv  \frac{\Delta V - d (\Delta V)/d\ln T}{\pi^2 g_* T^4/30},\qquad
\frac{\beta}{H}\equiv \frac{d(S_3/T)}{d\ln T}
\eeq
where 
$S_3$ is the action of the thermal bubble that determines the tunnelling rate per space-time volume as
$\Gamma \approx (S_3/2\pi T)^{3/2}  T^4 e^{-S_3/T}$ and $\Delta V$ is the potential difference between the two minima.
These quantities are evaluated at $T\approx T_f$, which is roughly determined as the temperature at which $S_3/T \approx 
4\ln M_{\rm Pl}/M_X\approx 142$.
Given that $S_3$ scales as $1/\beta_{\lambda_S}\propto g_X^4$, 
the result depends  strongly on $g_X$:
\begin{itemize}
\item For the critical value $g_X\approx 1.2$ one has 
$\alpha\approx 1$ in view of $T_f\approx 0.15 T_c$. Furthermore $\beta/H \approx 70$ such that $\Omega_{\rm peak} h^2\approx 2~10^{-11}$, which could be easily detected by planned experiments such as LISA.
Notice that $T_f \approx M_X/13\approx 110\GeV$ is larger than the DM freeze-out temperature, $T_{\rm fo}\approx M_X/25$
(such that DM freeze-out is negligibly affected by the phase transition) and is smaller than 
$T_c^{\rm SM}$ (such that also the SM scalar boson is involved in the first-order phase transition).

\item For $g_X\circa{>}1.2$ tunnelling is faster and consequently $T_f$ higher.
For example, for $g_X\approx 1.5$ one has $T_f\approx 0.5 T_c\approx 0.4 M_X\approx 800\GeV$ and $\alpha\approx 0.01$, $\beta/H\approx 200$ such that $\Omega_{\rm peak} h^2\approx 10^{-19}$, which could be detected by futuristic experiments.
The Higgs phase transition happens later independently of the dark phase transition.

\item For $g_X\circa{<}1.2$ thermal tunnelling is slower than the Hubble rate such that the universe would enter into an inflationary stage, which presumably ends when the temperature cools down to the $\SU(2)_X$ confinement scale $\Lambda_X$
after $N\approx \ln s_*/\Lambda_X \approx 8\pi^2/7g_X^2$ e-folds, next reheating the universe up to 
$g_* \pi^2 T_{\rm reh}^4/30 \approx \Delta V$ i.e.\ $T_{\rm reh}\approx M_X/9$.
The baryon asymmetry gets suppressed by a factor $\approx e^{-3N}$, such that the model is excluded when
this factor is smaller than the observed baryon asymmetry $\approx 10^{-9}$, provided that the baryon
asymmetry cannot be re-generated at the weak scale.
\end{itemize}
The critical value of $g_X$ could have an uncertainty of about $30\%$, given that higher order corrections
to the thermal potential are suppressed by $g_X/\pi$.




\bigskip

Finally note that the extended dark/electroweak phase transition also occurs for a more general situation with e.g.\ positive $H$ and $S$ squared mass terms typically smaller than the $v$ and $w$ symmetry breaking scales.

\bigskip

\section{Conclusions}\label{concl}
``Just so'' comparable small masses for the Higgs boson and for Dark Matter
(much smaller than the Planck scale) satisfy a reformulation of the naturalness concept,
modified by assuming that quadratic divergences should be ignored and thereby named `finite naturalness'.
Within this heretic point of view, it might be more satisfactory to
find a dynamical explanation of the smallness and of the proximity of these low-energy scales.
We here assumed that such masses vanish to start with, and that they originate from a physical mechanism
that occurs at such  energy scales.\footnote{This assumption is not demanded by  finite naturalness.
An alternative possibility compatible with this scenario is e.g.\ that
gravitational loops of unobservable intermediate-scale particles (as suggested to us by A. Arvinataki, S. Dimopoulos, S. Dubovsky) could generate small and comparable Higgs and scalar DM masses.}

This is achieved by a simple extension of the Standard Model, 
that contains just one extra scalar doublet under one extra $\SU(2)$ gauge group.
The extra vectors are automatically stable thermal DM candidates, just like the proton is
automatically stable within the SM.
The extra scalar doublet gives mass to such vectors
(because of their gauge interactions), to the SM Higgs boson
(because of a quartic coupling between them)
 and to itself (because its quartic couplings runs negative at low energy).
Thereby, all scales are related and exponentially  suppressed with respect to the Planck scale.

As function of only one free parameter the model predicts the properties of the
extra scalar (observable at collider experiments, see fig.\fig{FS}a), 
of Dark Matter (observable in direct and indirect detection experiments, see fig.\fig{FS}b).
The scalar potential of the model can be stable up to the Planck scale, even 
when the SM potential is unstable, namely for the present best-fit values of its parameters.
In cosmology, the model predicts a first order phase transition 
with emission of gravitational waves at a possibly detectable level.



\small

\paragraph{Acknowledgements}
We thank useful discussions with J.R.~Espinosa and M.H.G.~Tytgat. This work was supported by the ESF grant MTT8, by SF0690030s09 project, by the FNRS-FRS, by the IISN and by the Belgian Science Policy, IAP VII/37.

\bigskip

\newpage

\footnotesize
\begin{multicols}{2}

\end{multicols}

\end{document}